\documentclass[10pt,conference]{IEEEtran}
\IEEEoverridecommandlockouts
\usepackage{cite}
\usepackage{amsmath,amssymb,amsfonts}
\usepackage{algorithmic}
\usepackage{graphicx}
\usepackage{textcomp}
\usepackage{xcolor}
\def\BibTeX{{\rm B\kern-.05em{\sc i\kern-.025em b}\kern-.08em
    T\kern-.1667em\lower.7ex\hbox{E}\kern-.125emX}}
\begin{document}

\title{AC\textsuperscript{2} - Towards Understanding Architectural Changes in Python Projects
}

\author{\IEEEauthorblockN{A Eashaan Rao, Dheeraj Vagavolu, Sridhar Chimalakonda}
\IEEEauthorblockA{\textit{Research in Intelligent Software \& Human Analytics (RISHA) Lab}\\
Department of Computer Science \& Engineering \\
Indian Institute of Technology Tirupati India\\
\{cs19s501,cs17b028,ch\}@iittp.ac.in}
}

\maketitle

\begin{abstract}
Open source projects are adopting faster release cycles that reflect various changes. Therefore, comprehending the effects of these changes on software's architecture over the releases becomes necessary. However, it is challenging to keep architecture in-check and add new changes simultaneously for every release. To this end, we propose a visualization tool called AC\textsuperscript{2}, which allows its users to examine the alterations in the architecture at both higher and lower levels of abstraction for the python projects. AC\textsuperscript{2} uses call graphs and collaboration graphs to show the interaction between different architectural components. The tool provides four different views to see the architectural changes. The user can examine two releases at a time to comprehend the architectural changes between the releases. AC\textsuperscript{2} can support the maintainers and developers to observe changes in the project and its influence on the architecture, which allow them to see its increasing complexity over the releases at the component level. AC\textsuperscript{2} can be downloaded at \url{https://github.com/dheerajrox/AC2} and its demo can be seen at the website \url{https://dheerajrox.github.io/AC2doc} or on youtube \url{https://www.youtube.com/watch?v=GNrJfZ0RCVI} 
\end{abstract}

\begin{IEEEkeywords}
Software Architecture, Visualization Tool, Call Graphs, and Collaboration Graphs
\end{IEEEkeywords}

\section{Introduction}
Software architecture is considered as one of the critical artifacts that describe the interaction between different parts of the system, with well-thought design decisions \cite{breivold2010does}. However, as software evolves, its architecture becomes complex, leading to design erosion and high technical debts \cite{jansen2005software}. At the same time, many open source projects are embracing fast release engineering, resulting in frequent changes to the system \cite{khomh2015understanding}. These rapid modifications or enhancements at a short timescale put stress on the software system \cite{breivold2010does}. Therefore, understanding architectural changes as the software evolves allows teams to assess commercial and quality trade-offs and develop a better strategy to ensure system stability \cite{breivold2010does}. 


Researchers have proposed various tools to visualize architectural changes over multiple versions \cite{nam2018eva, mcnair2007visualizing,mcveigh2011evolve}. For instance, EVA \cite{nam2018eva} uses the ARCADE framework to show and compare architectural changes between multiple releases. YARN shows changes in architecture by showing its evolution using animation \cite{hindle2007yarn}. Although many architectural recovery techniques are proposed in the literature, given the source code of a project, understanding its architecture remains a great challenge, and to obtain its partial understanding, analysts break down the task to learn the relationships between different components \cite{ahn2009weighted}. One of the strategies they leverage is ``\textit{call and collaboration graphs}'' for representing the interaction between subroutines or class nodes \cite{russo2018profiling, ahn2009weighted}. These graphs help to demonstrate the dependency structure of software \cite{russo2018profiling}. Jaktman et al. employ call graphs to understand the signs of architectural erosion in the squid project \cite{jaktman1999structural}. Object-oriented notations such as \textit{classes}, \textit{collaborations}, \textit{interfaces}, and \textit{components} are used for architecture description \cite{garlan2002reconciling}. 
However, to the best of our knowledge, these graphs are not leveraged to visualize the software's architectural changes having multiple releases. Leveraging these graphs, we can have granular-level architecture and component-level interactions for better understanding.

In this paper, we propose a visualization tool called AC\textsuperscript{2} to comprehend the architecture changes in a software system over multiple releases for python projects. Although there are multiple ways to represent architecture, we limit our concept to denote architecture as a structure consisting of different components, and their interactions \cite{nakamura2005metrics} shown using call and collaboration graphs. AC\textsuperscript{2} uses call graphs that exhibit the interactions between subroutines, and collaboration graphs show the interaction among the classes.  AC\textsuperscript{2} compares architectural changes between different versions by visualizing it in a \textit{modular}, \textit{hierarchical}, and \textit{interactive} manner (see Section \ref{sec:ac_tool}).

\textbf{AC\textsuperscript{2}} could be distinguished from existing tools in terms of the following characteristics:
\begin{itemize}
    \item Architectural exploration is done using collaboration and call graphs.
    \item Visualize changes in the system at the directory, function, and class level for each release.
    \item These graphs show changes in the way a method or class interacts with other components over the releases.
\end{itemize}

The tool's main contribution is to facilitate understanding of software development based on its architectural changes allowing the developers and maintainers of the project to comprehend the state of the architecture.

\begin{figure*}
    \centering
    \includegraphics[scale=0.5]{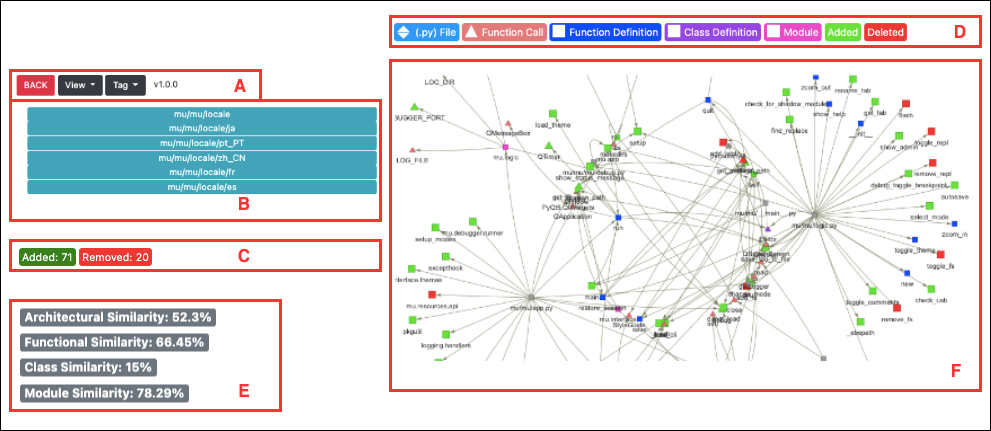}
    \caption{Snapshot of AC\textsuperscript{2} tool applied on a repository called \textit{mu-editor/mu}. In [A], ``View'' allows the user to see different views of architecture, and ``Tag'' will show the different releases of the repository. [B] shows the directory structure of the repository using which the user can traverse. [C] displays the added and removed components for a particular view. [D] displays the legend for the ``\textit{integrated view}'' (the legend will change according to the view) [E] shows the similarity between two versions based on the different components. [F] shows the ``\textit{integrated view}'' of 1.0.0 version of the \textit{mu-editor/mu} repository.}
    \label{fig:tool_view}
\end{figure*}

\section{AC\textsuperscript{2} Tool}
\label{sec:ac_tool}
This section describes the approach and architecture of AC\textsuperscript{2}. The tool supports GitHub’s open-source Python projects because it is a popular programming language used in GitHub\footnote{\url{https://octoverse.github.com/}}.

Figure \ref{fig:tool_view} shows a snapshot of AC\textsuperscript{2} depicting its various segments. Segments [A, B, and D] in figure \ref{fig:tool_view} lets users look at different views and versions, browse the repository directory, and a legend explaining the types of components present in a particular view. While segments [C and E] shows changes between two versions for a particular view, in terms of added and deleted components and similarity metrics (see Section \ref{sec: metrics}) and segment [E] shows one of the below mentioned views in figure \ref{fig:tool_view}, i.e., ``integrated view'' is shown for version 1.0.0 of repository \textit{mu-editor/mu\footnote{https://github.com/mu-editor/mu}}. AC\textsuperscript{2} displays four views of the architecture. Directory view showcases the architectural changes at a higher level of abstraction, while Call, Collaboration, and Integrated View display the lower-level details. The following list describes all these views and the rationale for selecting these views for AC\textsuperscript{2}: 
\begin{itemize}
    \item \textbf{Directory View}: It shows a hierarchical view describing the changes in the project's directories and file structure. In this view, we are referring to \textit{directories} and \textit{files} as architecture components at a higher level of abstraction. Understanding the higher-level changes in the project's organization provides an overview of the architecture since several directories and files are being added or removed over the releases. The directory's name also gives us the information related to big components in the architecture. For example, the directory name ``User Interface'' will tell us about the files and directories stored in it are related to the front end of the software.
    \item \textbf{Call Graph View}: It displays the interaction between subroutines by giving the information of callee and caller. In this view, we consider \textit{files,functions definition and function calls} as architectural components. The call graph generated in AC\textsuperscript{2} is for the specific directory which the user is exploring, not for the whole project. This modularization is necessary because understanding the call graph for the whole project is difficult, especially recognizing its different modules and functions.
    \item \textbf{Collaboration Graph View}: Similar to the call graph view, Collaboration graphs are generated for the respective directory on which the user wants to focus. In this view, \textit{files, class definitions}, and \textit{modules} are considered as architectural components. For the current version, we focus only on the classes present in the project code-base. The interaction between classes via inheritance and aggregation helps the user to understand the interdependence between different python files. Python file created by developers which act as modules is also incorporated in the view since they are not pre-defined libraries.
    \item \textbf{Integrated View}: It displays the combination of both call and collaboration graphs. This view gives an overall detailed view of the specific directory consists of components, i.e.,\textit{ files, function definition, function calls, class definition}, and \textit{modules}. This view can help the user to understand the interaction between classes and functions for the specific architecture module.
\end{itemize}



 
Showing changes in the interaction between components is crucial as it allows users to reflect on architectural changes at a deeper level. It tells us about the complexity with which architecture is evolving, which we try to exhibit using call and collaboration graphs. It provides users with a better understanding of the overall system.

\begin{figure}[ht]
    \centering
    \includegraphics[width=9cm, height=5cm]{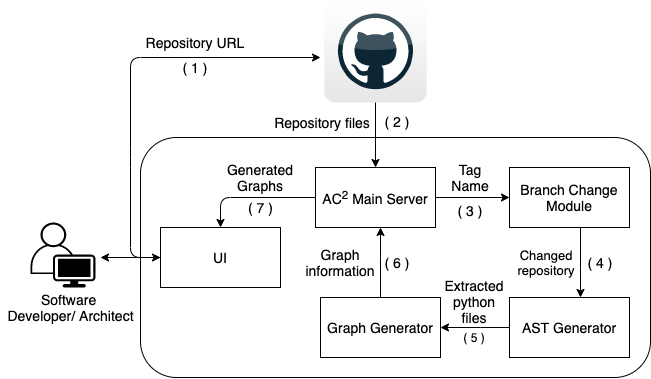}
    \caption{Architecture of AC\textsuperscript{2}}
    \label{fig:arch}
\end{figure}


\subsection{\textbf{Architecture}}
Figure \ref{fig:arch} describes the architecture of the AC\textsuperscript{2}. It contains the following components: 

a) \textbf{AC\textsuperscript{2} Main Server}: It is the control center of AC\textsuperscript{2}. It facilitates communication and interaction amongst all the other components. We used flask framework for building the server.
    
b) \textbf{Branch Change Module}:  This module helps the software navigate the various versions of the target repository and collects tags and release data of the repository directly from GitHub. This module is written in the shell scripting language. 
    
c) \textbf{AST Generator}: The AST generator works in sync with the ``branch change'' module and generates a mapping for each branch separately. It parses through the repository recursively to generate AST's for python files. 
    
d) \textbf{Graph Generator}: It uses the AST's and directory maps from the ``AST Generator'' module and extracts names and interaction of function and class definitions, function calls, modules, and files. It then creates graph data structures using the extracted information, which is then passed on to the UI.
    
e) \textbf{User Interface (UI)}: The UI's landing page allows users to enter a repository link and then passes the information to the main server. The user can then navigate the directory structure and observe four generated views as mentioned above. However, as the system evolves, so do the graphs representing them. Therefore, AC\textsuperscript{2} considers the project in a hierarchical manner, where different directories represent different modules, which in turn can divide into sub-modules. The tool uses \textit{vis.js library\footnote{\url{https://visjs.org/}}} to create \textit{network graphs}. It allows easy navigation and provides a detailed description of nodes. The user can scale the graph and stretch the edges to get a more unobstructed view. On clicking the nodes, it shows component type. AC\textsuperscript{2} comes with a multi-view front-end to compare two releases simultaneously. 


\subsection{\textbf{Metrics}}
\label{sec: metrics}
AC\textsuperscript{2} uses two metrics, i.e., cohesion values to identify how functions are tightly linked during releases and A2A metric for providing the degree of similarity between two releases which gives the better understanding of architectural changes. 
\begin{itemize}
    \item \textbf{Cohesion}: It measures the degree to which components are functionally related. It is calculated using LCOM4\footnote{\url{https://www.aivosto.com/project/help/pm-oo-cohesion.html}} metric for methods in the source code and stored the values in separate files for each version.  
    \item \textbf{Architecture-to-architecture(A2A)}: It gives the similarity measure of the component-level changes between the architectures between two releases. This metric is adopted from the study done by Le et al. \cite{le2015empirical}.
  \begin{align*}
        a2a(A_{i},A_{j})=(1- \frac{mto(A_{i},A_{j})}{aco(A_{i})+aco(A_{j})}) * 100 
   \end{align*}
     \begin{align*}
        mto(A_{i},A_{j})= remC(A_{i},A_{j}) + addC(A_{i},A_{j}) \\
            + remE(A_{i},A_{j}) + addE(A_{i},A_{j}) 
    \end{align*}
    \begin{align*}
        aco(A_{i}) =  addC(A_{\phi},A_{i}) + addE(A_{\phi},A_{i})
    \end{align*}    
    where \textit{a2a(A\textsubscript{i}, A\textsubscript{j})} gives the architecture similarity score between two releases. \textit{mto(A\textsubscript{i}, A\textsubscript{j})} provides number of operations needed for to change the architecture from A\textsubscript{i} to A\textsubscript{j}, while \textit{aco(A\textsubscript{i})} gives the number of operations to construct architecture \textit{A\textsubscript{i}} from scratch. \textit{addC} and \textit{remC} are the addition and removal operations of components while \textit{addE} and \textit{remE} are the operations related to the component's entities. We modified the original A2A metric in the context of AC\textsuperscript{2} tool, for instance, the \textit{architectural similarity} (Figure \ref{fig:tool_view} [E]) is calculated using directory view where directory is considered as a component and files as its entities. For \textit{functional, class and module similarity}, we considered call and collaboration view by modifying \textit{mto(A\textsubscript{i}, A\textsubscript{j})} to calculate only \textit{addC} and \textit{remC} as there is no entities present for function, classes and modules unlike directory which has a entities as files.
   
\end{itemize}

\section{Case Study}
We assessed AC\textsuperscript{2} on \textit{mu-editor/mu}, which is a python code-editor for beginners and it has 12 releases. For the analysis, we restricted our focus on the changes of a specific directory named ``mu\footnote{\url{https://github.com/mu-editor/mu/tree/master/mu}}'' which is the main functional directory of the project. The analysis is made based on four views in the AC\textsuperscript{2} tool and from the repository documentation. 

The first significant changes after the deployment reflect in v1.0.0.beta.15 release. We observed 4 new directories and 2 python files are added, and 1 python file is removed using directory view. The architectural similarity score between v.0.9.12 and v1.0.0.beta.15 is found to be 60.8\%. We observed major changes in methods using the call graph view with a functional similarity score of 71.88\%. However, the class similarity is gone down to 26.32\% because very few class definitions exist in the folder, and in v1.0.0.beta.15, a couple of classes are removed, which results in less similarity value.

After release v1.0.0.beta.15, remaining releases mostly focused on bug fixes and some improvements in existing code. Till the latest release, i.e., v1.0.3, no major architectural changes are observed, and only 43 new architectural components such as files, modules, and function definition are added, and 16 components are deleted. Using AC\textsuperscript{2}, we identified 106 changes in the ``mu/mu" folder from all 12 releases. On average, 7 components are being added, and 1 is removed in every release. We also observed that removed components are often being replaced with similar new functionalities. In addition to it, we also observed significant changes in the interaction between components over the releases. 

\section{Limitations}
\begin{itemize}
    \item Visualizing architecture using call and collaboration graphs provides a limited perspective. However, we believe that interactions shown through these graphs can help to understand architectural changes adequately. 
    \item Over the releases, the systems grow; therefore, generated graphs become large, and it requires time to comprehend the changes present in a release.
    \item Some components of the repository are not Python files, functions, classes, or modules. For those instances, the color component is shown as ``grey''. 
    \item The library used to show the different views, i.e., \textit{vis.js}, sometimes cannot render the graph properly due to which some components may not visible. However, re-running the tool for that repository will resolve the issue.
    \item In the current version in collaboration graphs, we have considered only class definition, and in the future version, we will include objects as well.
    \item In the \textit{call, collaboration} and \textit{integrated} view, it is not easy to visualize modularization of different files and its function and classes, which can allow a better understanding of these views. In subsequent studies, we plan to use better modularization techniques on these graphs.  
\end{itemize}

\section{Related Work}
A visualization tool called EVA explores the architecture and software changes using various architectural recovery techniques based on the ARCADE framework \cite{nam2018eva}. GASE uses architectural changes to compare two releases of a software \cite{holt1996gase}. 
YARN shows the cumulative changes in the software by showing architectural evolution through animation \cite{hindle2007yarn}. Le et al. proposed architecture recovery methods and presented an empirical study to comprehend architectural changes in multiple versions of open source projects \cite{le2015empirical}. BEAGLE provides web-based visualization and navigation to study architectural evolution and uses several metrics to understand evolution in a long-lived system \cite{tu2002integrated}. Evolve \cite{mcveigh2011evolve} captures incremental changes in the architecture using \textit{Backbone} an architecture description language. Motive \cite{mcnair2007visualizing} visualize the net changes in the software architecture to understand software evolution.


\section{Conclusion and Future Work}
AC\textsuperscript{2} is a visualization tool that shows architectural changes in python projects having multiple releases. It uses call and collaboration graphs to achieve this goal. It showcases the interaction between architectural components using \textit{directory, call graph, collaboration graph}, and \textit{integrated view}. AC\textsuperscript{2} compares two releases together, visualizing the changed components and the interaction between them. Metrics such as cohesion, architectural, functional, class, and module similarity allows the user to understand changes between two releases. Our work shows that these graphs can be applied to visually understand the architectural changes and potentially uncover useful insights from it. 

We plan to extend AC\textsuperscript{2} by improving its efficiency, adding support to more languages such as C++ and Java, and combining different visual techniques with better modularization to ease the navigation among the components. The inclusion of other metrics will help to identify technical debts giving a detailed and accurate analysis of the architectural changes. We also plan to utilize other architecture recovery options to view the architectural changes from multiple perspectives.
\bibliographystyle{IEEEtran}
\bibliography{references}

\end{document}